\documentclass[
    ,final            % use final for the camera ready runs
  ]
  {aipproc}

\layoutstyle{6x9}

\usepackage{graphicx}
\usepackage{natbib}
\bibpunct{(}{)}{,}{n}{}{,}

\begin{document}

\title[Temporal coherence length of light]{Temporal coherence length of light in semiclassical field theory models}

\classification{03.50.-z, 42.25.Kb}
\keywords      {coherence, emission model, light, photon, pulse, temporal coherence length, wave packet, wave-particle duality}

\author{Borys Jagielski\footnote{E-mail: borys.jagielski@fys.uio.no}}{
  address={Quantum Optics Laboratory, Department of Physics, University of Oslo, P.O. Box 1048 Blindern, 0316 Oslo, Norway}
}

\author{Johanne Lein}{
  address={Quantum Optics Laboratory, Department of Physics, University of Oslo, P.O. Box 1048 Blindern, 0316 Oslo, Norway}
}

\author{Arnt Inge Vistnes}{
  address={Quantum Optics Laboratory, Department of Physics, University of Oslo, P.O. Box 1048 Blindern, 0316 Oslo, Norway}
}

\begin{abstract}
The following work is motivated by the conceptual problems associated with the wave-particle duality and the notion of the photon. Two simple classical models for radiation from individual emitters are compared, one based on sines with random phasejumps, another based on pulse trains. The sum signal is calculated for a varying number of emitters. The focus lies on the final signal's statistical features quantified by means of the temporal coherence function and the temporal coherence length. We show how these features might be used to experimentally differentiate between the models. We also point to ambiguities in the definition of the temporal coherence length.
\end{abstract}

\maketitle

\section{Introduction}

Inquiries into the nature of light have a long tradition, and they have often propelled the scientific progress \citep{Sobel}. Still, no full consensus has been reached on how we conceptualize light as a physical phenomenon \citep{Roychoudhuri}. This lack of agreement lies of course at the heart of the wave-particle duality of light \citep{Milloni}. The two paradigms -- the wave picture based on classical field theory of electromagnetism, and the photon picture based on quantum mechanics -- work very well within their own domains, but they do conflict in a profound way when one tries to give an unambigous answer to the qualitative question ``What is light?''.

With a realistic attitude towards theoretical constructs, it is unsatisfactory to have to use different physical descriptions of light behaviour in one and the same experimental setup: For instance, the wave picture underlies propagation models, while we often switch to the particle picture when describing the detection process. We believe that one should strive for a greater accord by proposing new models of light or revising old ones. Of course, an ultimate resolution of the wave-particle duality is a very ambitious and long-term goal, but the research activity alone might enhance our understanding of light and possibly lead to novel predictions.

The two aforementioned domains of applicability are usually distinguished in terms of light intensity. When it is high (relative to $h\nu$ per unit time per unit area, $\nu$ being the central frequency of the quasi-monochromatic beam, and $h$ Planck's constant) the Maxwellian theory offers the best description; when the intensity is very low, we commonly employ the quantum framework with Fock states. Fock states, however, do not describe the propagation of individual photons, and the time evolution aspect is neglected.

Low-intensity light may be generated in two different ways: It can be obtained by starting out with high-intensity light which is subsequently attenuated outside of the source, or it may be produced directly by using very few emitting entities to begin with. It may well happen that the high-to-low transition proceeds along different lines in these two different contexts. In the present work we focus on the latter situation, i.e. we study the source intensity with hope of gaining new insights about the generated light.

We compare simple emission models numerically, with a time resolution that is unatainable experimentally. We examine how the superposition of their uncorrelated primitive signals gradually ``builds up'' the final signal when the number of emitters increases. The final signal is expected to depend on the primitive signal, i.e. on the emission model employed. Two models for the
primitive signal are studied so far. The signal could be a steady-state wave from each emitter \citep[Ch. 3.1]{Loudon2}, or it could consist of wave packages emitted once in a while from each emitter \citep{Janossy}. In this work we name these models ``continuous'' and ``pulsed'', respectively.

We argue that two final signals may look very similar in the high intensity regime, even though their underlying emission models are vastly different. The difference may reveal itself only when the number of emitters is very low. We show how temporal coherence length of the final signal may be used to discern between the emission models, that is, how the coherence length as a function of intensity depends on the model chosen.

\section{Methods}

We consider an ensemble of $n$ primitive optical oscillators. These could be atoms in a thermal source where each atom acts as an emitter. We model the resulting radiation classically by assuming that each oscillator emits an electromagnetic wave which is of somewhat regular character, but contains random features. The superposition of all waves yields the final stochastic optical signal. Notice that we consider a single component of the electromagnetic field only, and assume as a simplifying constraint that all radiation is being emitted in a single direction with the same polarization vector. The stochastic parameters do not change neither in time nor across the ensemble.

We compare two models, where each model is to be understood as a specific radiative mode of a primitive optical oscillator; in a given model all oscillators are characterized by the same mode. The first model (M1) is based on monochromatic waves which undergo random changes in phase~\citep[Ch. 3.1]{Loudon2}, matematically expressed as
\begin{equation}\label{eq:primsig1}
Y(t) = A \sin[\omega t + \Phi(t)],
\end{equation}
where $\Phi(t_{i} \leq t<t_{j})=\Phi_{i}$ is constant, $t_{i}$ and $t_{j}$ being the times of subsequent phase changes. The amplitude and wavelength are fixed for a single atom, but are normally distributed across the ensemble.

In the second model (M2) the primitive electromagnetic field consists of pulses modeled as sines enveloped in Gaussian functions~\citep{Janossy}\footnote{Note that Janossy assumed an instantaneous excitation with exponentially decaying pulses. We, however, delibaretely employ symmetric pulses.}: 
\begin{equation}\label{eq:primsig2}
Y(t)=\sum_{i=1}^{M}A_{i}e^{-[(t-t_{i})/C_{i}]^2}\sin(\omega_{i}t+\Phi_{i}),
\end{equation}
where $t_{i}$ is the central time of the $i$-th pulse, and $C_{i}$ determines the pulse damping. The amplitude, wavelength and damping are randomized from pulse to pulse according to a normal distribution. Both the random phase jumps in M1 and the emission times of pulses in M2 are Poisson distributed. 

The underlying physical picture in the case of M1 is that of each atom emitting a steady train of electromagnetic radiation until it suffers a collision. Assuming the collision time to be negligible, the collision causes an abrupt and random change in the phase of the radiation. In the case of M2 we suppose that each atom is excited at random times, and that each excitation results in an oscillation occurring within a limited time period (corresponding to the pulse width). Still, we stress that our focus lies on the statistical features of the emerging signals, not on the physical behaviour of the source atoms themselves. 

The signals have been generated numerically with routines programmed in C. Our central wavelength is 600 nm corresponding to a period\footnote{The time resolution in our model is 0.04~fs yielding 50 data points per period.} of 2~fs. With a total of $10^6$ data points, the simulated time interval will be 0.04 ns, corresponding to the light covering a distance in the order of milimeters.

\begin{figure}
\includegraphics[height=.5\textheight, width=1.0\textwidth]{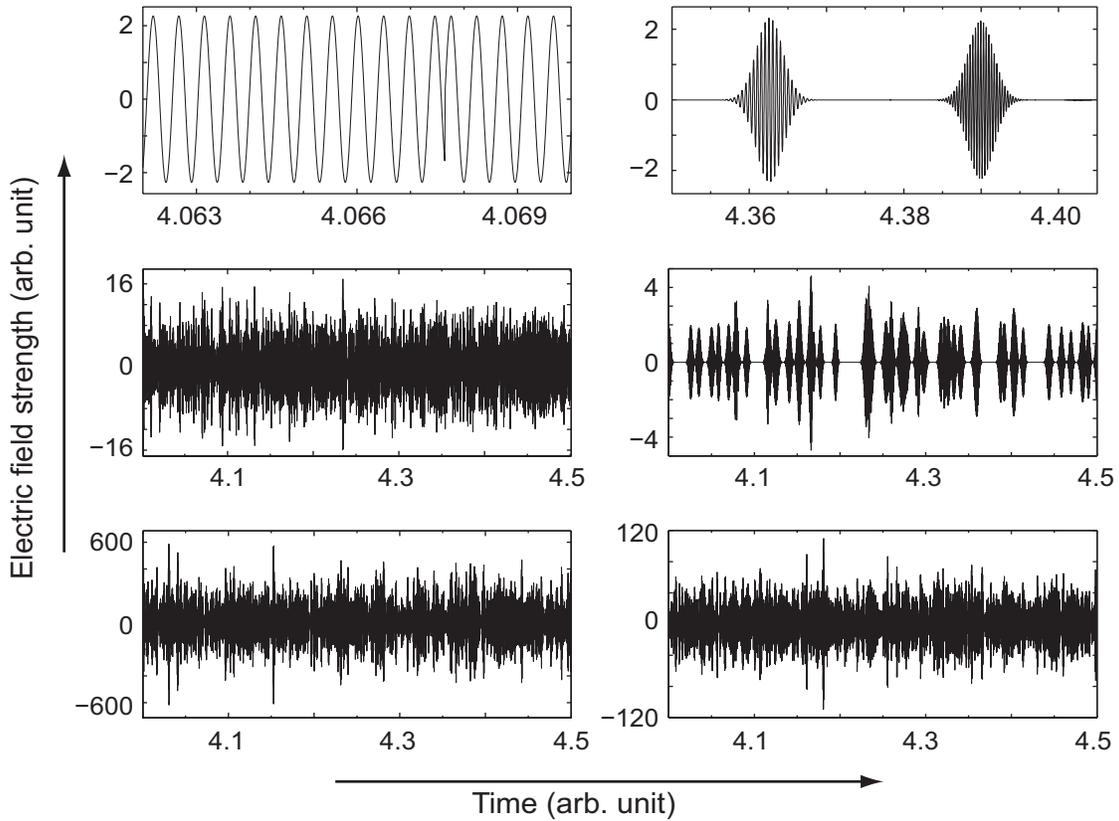}
\caption{Stochastic signals generated from the phase-jump model M1 (left column) and from the pulse model M2 (right column). Number of emitters increases row-wise and downwards: for the top row we have $n = 1$, for the middle row $n = 10$, and for the bottom row $n = 10000$. In the high-intensity regime (bottom) the signals are very similar, but when the number of emitters is decreased by three orders of magnitude (middle) the phasejump signal obviously preserves its continuous character, while the pulse signal ``disintegrates'' into separate pulses. Notice that the top row is scaled differently in the time direction than the rest of the figure.}
\label{fig:pulses_phasejumps}
\end{figure}

Figure \ref{fig:pulses_phasejumps} presents sample signals for M1 and M2 for $n=$1, 10, and $10^4$ emitters. For $n=10^4$, the signals look very similar, while for few emitters the difference in emission modes is visible. These features can be analyzed quantitatively using the first-order degree of temporal coherence (FODTC, or temporal coherence for short), as will be done in the next section. The FODTC for a random signal $E(t)$ is defined as:
\begin{equation}\label{eq:def_fodtc}
\gamma(\Delta t) = \frac{\langle E(t) E(t+\Delta t) \rangle}{\langle E(t) E(t) \rangle},
\end{equation}
with the angular brackets denoting time average\footnote{Which is equal to ensemble average given stationarity and ergodicity of the random process.}. The standard analytical definition \citep[Ch. 11.1B]{Saleh} of the temporal coherence length then reads:
\begin{equation}\label{eq:def_tcl}
l_c = c \int_{-\infty}^{\infty} |\gamma(\Delta t)|^2 d(\Delta t),
\end{equation}
with $c$ being the speed of light. This is the power-equivalent width of the FODTC. Alternatively one can approximate $l_c$ as Full Width at Half Maximum (FWHM) of $\gamma(\Delta t)$, but in the following we will point out that the latter definition, although convenient, is somewhat problematic.

\section{A sample of results and a short discussion}

\begin{figure}
\includegraphics{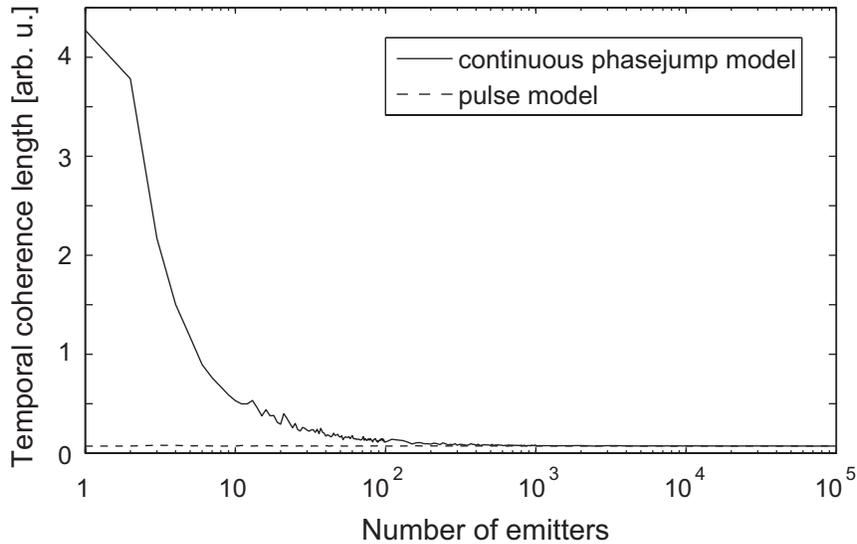}
\caption{Variation of the temporal coherence length $l_c$ for the two models as a function of the number of emitters (i.e. the intensity). The signals were 0.04~ns long ($10^6$ datapoints in total), and the FODTC was calculated up to $\Delta t = \pm$800~fs. $l_c$ was calculated from Eq.~\eqref{eq:def_tcl} with the integration performed over the whole integral. The mean period was 2~fs, and the mean pulse length was 50 periods. The periods, the amplitudes and the pulse lengths were normally distributed with standard deviations equal to 0.1 of the mean values. The Poisson rate for the random events (phasejumps in M1 or pulse emission in M2) was 0.0001 yielding approximately 100 random events per primitive signal.}
\label{fig:noe_dependency}
\end{figure}

Figure \ref{fig:noe_dependency} shows how the temporal coherence length, calculated from Eq.~\eqref{eq:def_tcl}, changes for the two models when the number of emitters varies from 1 to $10^5$. The stochastic parameters are fixed and analogous in both cases; their numerical values are given in the figure caption. We observe that while in the high-intensity regime $l_c$ converge rending the models indistinguishable, in the low-intensity regime $l_c$ increases considerably in M1 while it remains constant in M2. In other words, the models predict different coherence properties for the low-intensity light in the two cases\footnote{The constancy of $l_c$ in M2 is not a feature inherent to the pulse-based model. Another reasonable choice of the stochastic parameters could lead to some variation of the coherence length within that model as well.}.

\begin{figure}
\includegraphics{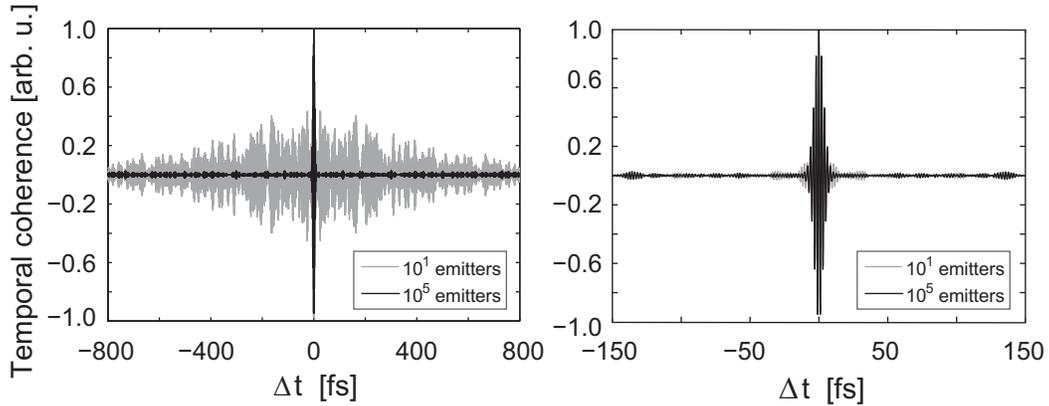}
\caption{FODTC's determined for the final signals generated using the phasejump model (\textbf{left}) and the pulse model (\textbf{right}), with $n=10$ (grey) and $n=10^5$ (thick black). While in the pulse model the FODTC's remain approximately the same when the number of emitters varies (and thus the two plots overlap rather tightly), in the phasejump model the temporal coherence function changes. This leads to the variation of the temporal coherence length as shown in Fig. \ref{fig:noe_dependency}.}
\label{fig:temporal_coherence}
\end{figure}

The qualitative reason for the different $l_c$ in M1 can be understood by considering Fig. \ref{fig:temporal_coherence} which shows the FODTC's of the final signals from the two models, with $n=10$ and with $n=10^5$. M2 always yields a well-defined central peak. Its shape and width do not change with the emitter number due to the fact that in any single pulse train the pulses are already independent. However, for M1 the situation is different. Here the underlying primitive signal is a continuous sine. Without any random phasejumps the coherence length would be infinite. The introduction of these random events serves to reduce $l_c$ by introducing more and more irregularities into the final signal. Thus, to begin with, the envelope of the FODTC decreases slowly, but eventually the central peak emerges. The ``fluctuations'' outside of it indicate the final signal is slightly correlated beyond its characteristic temporal coherence length. These secondary correlations diminish (relative to the central peak) as the intensity is increased even further. The fluctuations could be quantified, and their properties might be pursued experimentally through noise measurements beyond the central interference peak. Thus, noise measumerents might give another tool in discriminating between various models of light.

The left part of Fig. \ref{fig:temporal_coherence} also illustrates some ambiguities associated with the definition of the temporal coherence length. In our work this quantity has been calculated using the theoretical definition given in Eq.~\eqref{eq:def_tcl}. $l_c$ determined in this way subsumes both the total size of the central peak and the strength of the secondary correlations. However, the practical measurements of $l_c$ are concerned more with the FWHM of the FODTC. The secondary correlations will often be negligible, and then Eq.~\eqref{eq:def_tcl} will give a measure of the peak only, as required by the standard experimental approach. Nevertheless, already in our simple model, M1, the FWHM is difficult to ascertain when the intensity is low. This suggests that in some, not necessarily contrived, cases the FWHM-based definition and the definition based on the power-equivalent width are not as equivalent as one might implicitly assume. Such lack of equivalence might be of importance in some aspects of theoretical work, or when estimating temporal coherence length in delicate experimental situations.

We stress that in this preliminary work we do not aim at modeling any specific light sources exactly. Therefore the time scales involved, and the chosen values of the simulation parameters, are arbitrary, but not unrealistic. Here, our main goal is to consider the role of coherence analysis in practical discrimination between a continuous and a pulse-based classical model of radiation.

\section{Summary, conclusion, outlooks}
Our work so far indicates a direction our study of semiclassical field theory models may take. Above we have succinctly discussed two such models, and we have shown how the temporal coherence length might be used in the analysis. The variation of $l_c$ as a function of the number of emitters demonstrates that the transition from the high-intensity to the low-intensity regime should be understood differently in these models. Also, we have argued that there are ambiguities in connection with the definition(s) of the temporal coherence length which must be clarified when studying the transition region in more detail.

As explained in the introduction, we are motivated by the conceptual conflicts associated with the wave-particle duality and with the notion of the photon. We believe that in this context light coherence is a highly interesting property: It can be measured and analyzed in both frameworks \citep{MandelWolf}, although from the qualitative point of view its employment seems to be more rational in semiclassical models. One of the main reasons is that when discussing fluctuation correlations, we are in fact discussing phase correlations, and attempts at constructing phase operator for the quantized field are problem-ridden \citep[Ch. 10.7]{MandelWolf}. Thus, coherence may possibly serve not only as a statistical tool (as it has done above), but also as more general means to bring the two frameworks closer together.

It may be said that at the current stage our analysis is of more statistical than physical interest. Admittedly we have chosen not to analyze spontaneous radiation processes using quantum-mechanical principles in the spirit of Dicke \citep{Dicke}, nor to show how the coherence originates and grows within a source when these processes and principles are taken into consideration. However, in the long term the two approaches could be merged, i.e. the statistical analysis could be performed on models that are physically well-grounded, with focus still being kept on the differences between the high- and low-intensity domains. Already at this stage we have shown that it is principially possible to explore different emission models experimentally through coherence measurements. We feel that there has been little attention paid in the litterature to the possible dependencies of the coherence length on the intensity or other source parameters. Examining these connections may lead to useful insights into the nature of light.

In further work we wish to employ additional emission models, both continuous (e.g. with a random frequency walk within a primitive signal) and pulsed (e.g. with asymmetric pulses). We aim at a numerical survey of how the temporal coherence length changes when one varies different stochastic parameters. Spectral profiles will be taken into considerations, as there exists a strong link between them and the temporal coherence functions, formalized by the Wiener-Khinchin theorem \citep[Ch. 2.4.1]{MandelWolf}. However, the theorem presupposes stationarity of the random process, so it will be worthwhile to examine the stationarity demand in the low-intensity regime. Our models will also be substantiated by drawing analogies with the well-known broadening phenomena. The models might be further extended by including polarization effects, moving to two or three dimensions, or introducing some forms of dependency between the primitive signals (reflecting interactions between the emitters).

These studies will hopefully lead to more concrete experimental proposals on the one hand, and on the other hand they may help to evaluate the suitability of a ``wave packet'' as a classical conceptual equivalent of the photon when analyzing light behaviour.

\end{document}